\documentclass[12pt]{article}
\usepackage{amsmath,amssymb,amsthm}
\usepackage{enumerate}
\usepackage{latexsym}
\usepackage{pictex}
\usepackage[mathscr]{eucal}
\usepackage{graphicx,color,psfrag}

\newtheorem{theorem}{Theorem}[section]
\newtheorem{proposition}[theorem]{Proposition}

\newtheorem{lem}[theorem]{Lemma}
\newtheorem{rem}{Remark}[section]
\newtheorem{rems}{Remarks}[section]

\renewcommand{\epsilon}{\varepsilon}

\input{tcilatex}

\newcommand{\PRE}{{\it Phys. Rev. E} }

\newcommand{\JPA}{{\it J. Phys. A: Math. Theor.} }
\newcommand{\JMP}{{\it J. Math. Phys.} }
\newcommand{\jpa}{{\it J. Phys. A: Math. Gen.} }

\newcommand{\LMP}{{\it Lett. Math. Phys.} }

\newcommand{\MC}{{\it Math. Comp.} }
\newcommand{\JMA}{{\it J. Math. Anal.} }

\newcommand{\PAMS}{{\it Proc. Am. Math. Soc. }}

\newcommand{\RMP}{{\it Rev. Math. Phys.}}

\newcommand{\MPAG}{{\it Math. Phy. Ana. Geom.}}
\newcommand{\JSP}{{\it Jour. Stat. Phy.}}
\newcommand{\JFA}{{\it J. Funct. Anal}}
\begin{document}
\title{A quantum waveguide with Aharonov-Bohm Magnetic Field}
\author{H. Najar
\thanks{D\'epartement de Math\'ematiques, Facult\'{e} des Sciences de Moanstir. Avenue de
l'environnement 5019 Monastir -TUNISIE.\newline Laboratoire de
recherche: Alg\`{e}bre Géométrie et Th\'{e}orie Spectrale :
LR11ES53}\and M. Raissi$^{\ast}$}
\date{}
\maketitle

\begin{abstract}
 In a previous study \cite{n} we investigate
the bound states of the Hamiltonian describing a quantum particle
living on three dimensional straight strip of width $d$. We impose
the Neumann boundary condition on a disc window of radius $a$ and
Dirichlet boundary conditions on the remained part of the boundary
of the strip. We proved that such system exhibits discrete
eigenvalues below the essential spectrum for any $a>0$. In the
present work we study the effect of the presence of a magnetic field
of Aharonov-Bohm type on this system. Precisely we prove that in the
presence of such field there is some critical values of $a_0>0$, for
which we have absence of the discrete spectrum for $\displaystyle
0<\frac{a}{d}<a_0$. We give a sufficient condition for the existence
of discrete eigenvalues.
\end{abstract}

\baselineskip=20pt \setcounter{section}{0}
\renewcommand{\theequation}{\arabic{section}.\arabic{equation}}

\noindent

\section{Introduction}
 The task of finding eigenmodes
$(E_n,f_n), n=1,2,...$ of the Laplacian in the two-dimensional (2D)
and three-dimensional (3D) domain $\Omega$ with mixed Dirichlet
\begin{equation}\label{Dirichlet1}
\left.f_n(r)\right|_{\partial\Omega_D}=0
\end{equation}
and Neumann
\begin{equation}\label{Neumann1}
\left.{\bf n}{\nabla}f_n(r)\right|_{\partial\Omega_N}=0,
\end{equation}
boundary conditions on its confining surface (for 3D) or line (for
2D) $\partial\Omega=\partial\Omega_D\cup\partial\Omega_N$ (where
$r=(x,y)$ for two dimensions or $r=(x,y,z)$ for three dimensions and
$\bf n$ is a unit normal vector to $\partial\Omega$)
\cite{Borisov2,Dowker1,Jakobson1,Levitin1,Najar1,Seeley1} is
commonly referred to as Zaremba problem \cite{Zaremba1}, it is a
known mathematical problem science. Apart from the purely
mathematical interest, an analysis of such solutions is of a large
practical significance as they describe miscellaneous physical
systems. For example, the temperature $T$ of the solid ball floating
in the ice water  obeys the Neumann condition on the part of the
boundary which is in the air while the underwater section of the
body imposes on $T$ the Dirichlet demand \cite{Dowker1}. Mixed
boundary conditions were applied for the study of the spectral
properties of the quantized barrier billiards and of the ray
splitting in a variety of physical situations. The problem of the
Neumann disc in the Dirichlet plane emerges naturally in
electrostatics \cite{Jackson1}. In the limit of the vanishing
Dirichlet part of the border, the reciprocal of the first eigenvalue
describes the mean first passage time of Brownian motion to
$\partial\Omega_D$. In cellular biology, the study of the diffusive
motion of ions or molecules in neurobiological microstructures
essentially employs the combination of these two types of the
boundary conditions on the different parts of the confinement
\cite{Schuss1}.

One class of Zaremba geometries, that recently received a lot of
attention from mathematicians and physicists, are 2D and 3D straight
and bent quantum wave guides
\cite{n,Borisov10,bulla,Davies1,Dittrich1,exner1,Krejcirik1,Olendski1,Najar1}.
In particular, the conditions for the existence of the bound states
and resonances in such classically unbounded system were considered
for the miscellaneous permutations of the Dirichlet and Neumann
domains \cite{Dittrich1,exner1,Olendski2}. Bound states, lying below
the essential spectrum of the corresponding straight part, were
predicted to exist for the curved 2D channel if its inner and outer
interfaces support the  Dirichlet and Neumann requirements,
respectively, and not for the opposite configuration
\cite{Dittrich2,Krejcirik1,Olendski1}. This was an extension of the
previous theoretical studies of the existence of the bound states
for the pure Dirichlet bent wave guide \cite{Exner2} that were
confirmed experimentally. Also, for the 2D straight Dirichlet wave
guide the existence of the bound state below the essential spectrum
was predicted when the Neumann window is placed on its confining
surface \cite{bulla,exner1}. From practical point of view, such
configuration can be realized in the form of the two window-coupled
semiconductor channels of equal widths \cite{exner1,Exner6} whose
experimental creation and study have been made possible due to the
advances of the modern growth nanotechnologies. The number of the
bound states increases with the window length $a$ and their energies
are monotonically decreasing functions of $a$ \cite{Borisov10}. In
particular, for small values of $a$ the eigenvalue emerges from the
continuous spectrum proportionally to $a^4$. The asymptotical
estimate for small $a$ were established in \cite{Exner3}. The
asymptotic expansions of the emerging eigenvalue for small $a$ was
constructed formally in \cite{IY}, while the rigorous results were
obtained in \cite{RR}. Recently, this result was extended to the
case of the 3D spatial Dirichlet duct with circular Neumann disc
\cite{n} for which the bound state existence was confirmed, the
number of discrete eigenvalues as a function of the disc radius $a$
was evaluated and their asymptotics for the large $a$ was given. As
mentioned above, such Zaremba configuration is indispensable for the
investigation of the electrostatic phenomena \cite{Jackson1}.
Similar to the 2D case, it can be also considered as  the equal
widths limit of the two 3D coupled Dirichlet ducts of different
widths in general with the window in their common boundary
\cite{Borisov12,Exner6}.

The study of quantum waves on quantum waveguide has gained much
interest and has been intensively studied during the last years for
their important physical consequences. The main reason is that they
represent an interesting physical effect with important applications
in nanophysical devices, but also in flat electromagnetic waveguide.
\newline Exner et al. have done seminal works in this field. They
obtained results in different contexts, we quote
\cite{exner4,Exner2,exner1,Exner3}. Also in
\cite{Stokel,speis1,naj6} research has been conducted in this area;
the first is about the discrete case and the two others for deals
with the random quantum waveguide.

It should be noticed that the spectral properties essentially
depends on the geometry of the waveguide. In particular, the planar
waveguide with corners (small angle limit) \cite{mn} and the
existence of a bound states, which was induced by curvature
\cite{bulla,Dittrich1,exner4,Exner2} or by coupling of straight
waveguide through windows \cite{Exner2}.

On the other hand, the results on the discrete spectrum of a
magnetic Schr\"{o}dinger operator in waveguide-type domains are
scarce. A planar quantum waveguide with constant magnetic field and
a potential well is studied in \cite{PPB}, where it was proved that
if the potential well is purely attractive, then at least one bound
state will appear for any value of the magnetic field. Stability of
the bottom of the spectrum of a magnetic Schr\"{o}dinger operator
was also studied in \cite{EK,TW}. Moreover their influence on the
Dirichlet-Neumann structures was analyzed in
\cite{Borisov13,Olendski2}, the first dealing with a smooth
compactly supported field as well as with the Aharonov-Bohm field in
a two dimensional strip and the second with perpendicular
homogeneous field in the quasi dimensional.

Despite numerous investigations of quantum waveguides during last
few years, many questions remain to be answered. This concerns, in
particular, effects of external fields. Most attention has been paid
to magnetic fields, either perpendicular to the waveguides plane or
threaded through the tube. While the influence of the Aharonov-Bohm
field alone remained mostly untreated.

In their celebrated 1959 paper \cite{AB} Aharonov and Bohm pointed
out that while the fundamental equations of motion in classical
mechanics can always be expressed in terms of field alone, in
quantum mechanics the canonical formalism is necessary, and as a
result, the potentials cannot be eliminated from the basic
equations. They proposed several experiments and showed that an
electron can be influenced by the potentials even if no field acts
upon it. More precisely, in a field-free multiply-connected region
of space, the physical properties of a system depend on the
potentials through the gauge-invariant quantity $\oint \mathbf{A}
dl$, where $\mathbf{A}$ represents the vector potential. Moreover,
the Aharonov-Bohm effect only exists in {\textbf{the
multiply-connected region of space}. The Aharonov-Bohm experiment
allows in principle to measure the decomposition into homotopy
classes of the quantum mechanical propagator.

A possible next generalization are waveguides with combined
Dirichlet and Neumann boundary conditions on different parts of the
boundary with an Aharonov-Bohm magnetic field with the flux
$2\pi\alpha$. The presence of different boundary conditions and such
field also gives rise to nontrivial properties like the existence of
bound states. This question is the main object of the paper. The
rest of the paper is organized as follows, in Section 2, we define
the model and recall some known results. In section 3, we present
the main result of this note followed by a discussion. Section 4 is
devoted for numerical computations.

\section{The model}
The system we are going to study is given in Fig 1. We consider a
Schr\"odinger particle whose motion is confined to a pair of
parallel plans of width $d$. For simplicity, we assume that they are
placed at $z=0$ and $z=d$. We shall denote this configuration space
by $\Omega_0$
\[
\Omega_0=\Bbb{R}^2\times [0,d].
\]
Let $\gamma(a)$ be a disc of radius $a$, without loss of generality
we assume that the center of $\gamma(a)$ is the point $(0,0,0)$;
\begin{equation}
\gamma(a)=\{(x,y,0)\in \Bbb{R}^3;\ x^2+y^2\leq a^2\}.
\end{equation}
We set $\Gamma=\partial\Omega_0\diagdown \gamma(a)$. We consider
Dirichlet boundary condition on $\Gamma$ and Neumann boundary
condition on $\gamma(a)$.

\begin{figure}
\centering
\includegraphics[width=1\columnwidth]{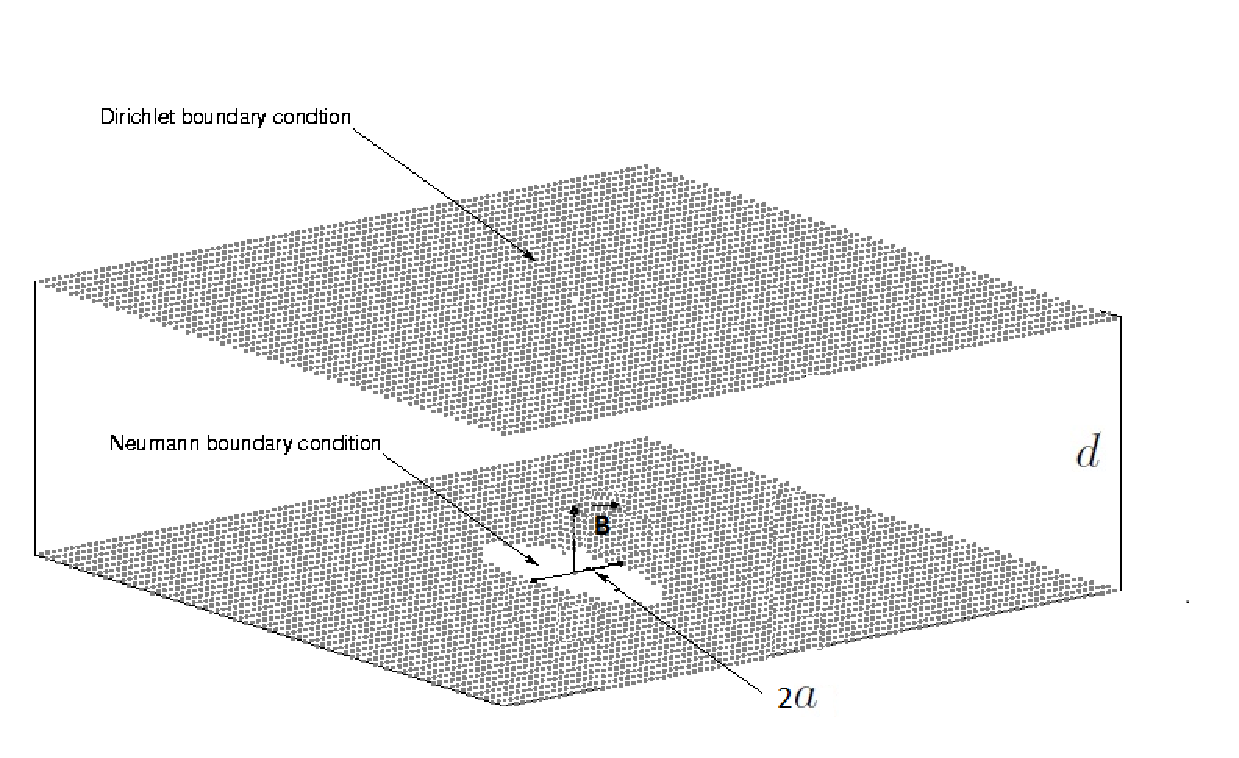}
\caption{\label{fig1} The waveguide with a disc window and  two
different boundaries conditions.}
\end{figure}



\newpage
\subsection{The Hamiltonian}
 Let be the multiply-connected region $\Omega=\{(x,y,z)\in\Omega_0; \quad
 x^2+y^2>0\}$. Let us define now the self-adjoint operator on
$\mathrm{L}^2(\Omega)$ corresponding to the particle Hamiltonian
$H$. For this we use the mean of Friedrichs extension theorem.
Precisely, let $H_{AB}$ be
 the Aharonov-Bohm Schr\"{o}dinger operator in
$\mathrm{L}^2(\Omega)$, defined initially on the domain
$\mathrm{C}_0^{\infty}(\Omega)$, and
 given by the expression
\begin{equation}\label{1}
  H_{AB} = (i\nabla+\mathbf{A})^2,
\end{equation}
where $\mathbf{A}$ is a magnetic vector potential for the
Aharonov-Bohm magnetic field $\mathbf{B}$, and given by
\begin{equation}\label{A}
  \mathbf{A}(x,y,z)=(A_1,A_2,A_3)=\alpha\biggl(
  \frac{y}{x^2+y^2},\frac{-x}{x^2+y^2},0\biggr),\quad
  \alpha\in(0,1).
\end{equation}
The magnetic field $\mathbf{B}:\mathbb{R}^3\rightarrow \mathbb{R}^3$
is given by
\begin{equation*}
    \mathbf{B}(x,y,z)=curl\mathbf{A}=0,
\end{equation*}
outside the $z$-axis and
\begin{equation}
\int_{\varrho}\mathbf{A}=2\pi\alpha,
\end{equation}
where $\varrho$ is a properly oriented closed curve which encloses
the $z$-axis. It can be shown that $H_{AB}$ has a four-parameter
family of self-adjoint extensions which is constructed by means of
von Neumann's extension theory \cite{adami,bra,dab}. Here we are
only interested in the Friedrichs extension of $H_{AB}$ on
$\mathrm{L}^2(\Omega)$ which we construct now by
means of quadratic forms.\\
For $\mathbf{A}=(A_1, A_2, 0)$ in \eqref{A}, we observe that
$A_1,A_2\in\mathrm{L}_{loc}^{\infty}(\Omega)$. Let
$$\Omega_n=B(0,n)\times[0,d]\setminus \left(B(0,1/n)\times[0,d]\right),\quad n\geq2,$$ where $B(0,r)$
denotes the disk with center $0$ and radius $r$. We define on
$\mathrm{L}^2(\Omega_n)$ (for each $n\geq2$) the sesquilinear form
\begin{eqnarray*}
  q_n[u,v] &=& \int_{\Omega_n}\biggl( i\frac{\partial u}{\partial x}+ A_1 u\biggr) \overline{\biggl(i\frac{\partial v}{\partial x}+ A_1 v \biggr)}
  + \int_{\Omega_n}\biggl( i\frac{\partial u}{\partial y}+ A_2 u\biggr) \overline{\biggl(i\frac{\partial v}{\partial y}+ A_2 v
  \biggr)}\\
  && +\int_{\Omega_n} \biggl( i\frac{\partial u}{\partial z}+ A_3 u\biggr) \overline{\biggl(i\frac{\partial v}{\partial z}+ A_3 v
  \biggr)},
\end{eqnarray*}
on the domain $$\mathrm{Q}(q_n)=\{ u\in \mathrm{H}^1(\Omega_n);\quad
u\lceil\Gamma_n=0\},$$ where $\displaystyle
\mathrm{H}^1(\Omega_n)=\{ u\in\mathrm{L}^2(\Omega_n)|\nabla
u\in\mathrm{L}^2(\Omega_n)\}$ is the standard Sobolev space,
$\Gamma_n=\partial\Omega_n\setminus\gamma(a)$ and we denote by
$u\lceil\Gamma_n$, the trace of the function $u$ on $\Gamma_n$. The
form $q_n$ is closed, since
$A_1,A_2\in\mathrm{L}^{\infty}(\Omega_n)$.\\Let the quadratic form
$q$
\begin{eqnarray*}
  q[u,v] &=& q_n[u,v] \quad\textrm{ if } u, v \in\mathrm{Q}(q_n), \\
  \mathrm{Q}(q) &=& \cup_n\mathrm{Q}(q_n).
\end{eqnarray*}
The form $q$ is symmetric and semi-bounded.
\begin{lem}\label{lem}
The form $q$ is closable.
\end{lem}
\noindent\textbf{Proof.} The form $q$ is closable if and only if any
sequence, $u_n\in\mathrm{Q}(q)$, such that
\begin{equation}\label{ph}
   \lim_{n\rightarrow\infty}\parallel u_n\parallel=
   0\quad\textrm{and }\quad\lim_{m,n\rightarrow\infty}q[u_n-u_m]=0,
\end{equation}
satisfies $\lim_{n\rightarrow\infty}q[u_n]=0$. \\Observe that
\eqref{ph} implies
\begin{equation}\label{4}
C:=\sup_nq[u_n]^{1/2}<\infty.
\end{equation}
Let $(u_n)_n\in\mathrm{Q}(q)$ be a sequence such that \eqref{ph} is
satisfied, then we take $\epsilon>0$ and choose $n_0$ such that
\begin{equation}\label{5}
q[u_n-u_m]\leq \epsilon \quad\textrm{for}\quad n,m\geq n_0,
\end{equation}
and
\begin{equation}
\parallel u_n\parallel\leq
\epsilon\quad\textrm{for}\quad n\geq n_0.
\end{equation}
Set, moreover, $K=\Omega_{n_0}\subset\Omega$ such that the support
of
 $u_{n_0}$ is included in $K$. In view of \eqref{ph} it follows
that
\begin{equation}\label{6}
\int_K\mid  (i\nabla+\mathbf{A})(u_n-u_m)\mid^2 \leq
q[u_n-u_m]\longrightarrow 0 \quad\textrm{as}\quad
n,m\rightarrow\infty,
\end{equation}
\begin{equation}\label{7}
\int_K\mid u_n\mid^2 \longrightarrow 0 \quad\textrm{as}\quad
n\rightarrow\infty.
\end{equation}
 $\mathbf{A}$ is bounded on $K$ as
 \begin{equation}
\parallel \mathbf{A}\parallel\leq \mid K\mid n_0,\quad\textrm{where $\mid K\mid$ denotes the lebesgue measure of $K$}
 \end{equation}
  we obtain that
\begin{equation}\label{8}
\int_K\mid \mathbf{A} u_n\mid^2 \longrightarrow 0
\quad\textrm{as}\quad n\rightarrow\infty.
\end{equation}
Since the norm in $\mathrm{L}^2$ is 1-Lipschitz, then
\begin{center}
    $\biggl|\biggl(\displaystyle\int_K\mid \mathbf{A}(u_n-u_m)\mid^2dx\biggr)^{1/2}-\biggl(\int_K\mid
    \nabla(u_n-u_m)\mid^2dx\biggl)^{1/2}\biggr|$
\end{center}
\begin{center}
   $\leq \biggl(\displaystyle\int_K\mid (i\nabla +\mathbf{A})(u_n-u_m)\mid^2dx\biggr)^{1/2}.$
\end{center}

\noindent According to \eqref{7}, the first term on the left side of
the latter tends to zero as $n,m\rightarrow\infty$ and, due to
\eqref{5}, the same holds for the right side. Thus,
\begin{eqnarray}\label{itoil}
  \int_K \mid u_n-u_m\mid^2+\mid\nabla( u_n-u_m)\mid^2 &\longrightarrow& 0 \quad\textrm{as}\quad
n,m\rightarrow\infty.
\end{eqnarray}
Since the form of the classical Dirichlet-Neumann Laplacian in
$\mathrm{Q}(q_{n_0})$ is closable, it follows from (\ref{itoil}) in
conjunction with \eqref{6} that
\begin{equation}\label{9}
\int_K\mid\nabla u_n\mid^2 \rightarrow0,\quad\int_K\mid u_n\mid^2
\rightarrow0,\quad\textrm{as}\quad n\rightarrow\infty.
\end{equation}
Let us consider the following quadratic form
\begin{equation}\label{10}
q[u_n]=q[u_n, u_n-u_{n_0}]+q[u_n,u_{n_0}]\\
\leq q[u_n]^{1/2}q[u_n-u_{n_0}]^{1/2}+\mid q[u_n-u_{n_0}]\mid.
\end{equation}
It follows from \eqref{4} and \eqref{5} that
\begin{equation}\label{11}
q[u_n]^{1/2}q[u_n-u_{n_0}]^{1/2}\leq C \epsilon^{1/2}
\quad\textrm{when} \quad n\geq n_0.
\end{equation}
Since $\mathbf{A}$ is bounded on $K$ we infer from \eqref{8} and
\eqref{9} that
\begin{equation}\label{12}
q[u_n,u_{n_0}]=\int_K
(i\nabla+\mathbf{A})u_n\overline{(i\nabla+\mathbf{A})u_{n_0}}
\rightarrow0\quad\textrm{as}\quad n\rightarrow\infty.
\end{equation}
Using \eqref{11} and \eqref{12} in \eqref{10} shows that
$\lim_{n\rightarrow\infty}q[u_n]=0$, this ends the proof of the
lemma \ref{lem}.
 $\blacksquare$\\

We denote the closure of $q$ by $\overline{q}$ and the associated
semi-bounded self-adjoint operator is the Friedrichs extension of
$H_{AB}$ is denoted by $H$ and its domain by $D(\Omega)$. It is the
hamiltonian describing our system. We conclude that the domain
$D(\Omega)$ of $H$ is
\begin{eqnarray*}
   D(\Omega)&=& \{ u\in\mathrm{H}^1(\Omega);\quad (i\nabla+\mathbf{A})^2
   u\in\mathrm{L}^2(\Omega),u\lceil\Gamma=0,\nu.(i\nabla+\mathbf{A})u\lceil\gamma(a)=0\},
\end{eqnarray*}
 where $\nu$ the normal vector and
\begin{equation}\label{13}
Hu=(i\nabla+\mathbf{A})^2 u, \quad \forall u\in D(\Omega).
\end{equation}

\subsection{Some known facts}
Let's start this subsection by recalling that in the particular case
when $a = 0$, we get $H^0$, the magnetic Dirichlet Laplacian, and
when $a =+\infty$ we get $H^{\infty}$, the magnetic
Dirichlet-Neumann Laplacian.
\begin{proposition}\label{th}
The spectrum of $H^0$ is $[(\frac{\pi}{d})^2, +\infty[,$ and the
spectrum of $H^{\infty}$ coincides with $[(\frac{\pi}{2d})^2,
+\infty[.$
\end{proposition}
\noindent\textbf{Proof.} We have
\begin{eqnarray*}
  H &=& (i\nabla+\widetilde{\mathbf{A}})^2\otimes
  I\oplus I \otimes(-\Delta_{[0,d]}),\quad
  \textrm{on}\quad\mathrm{L}^2(\mathbb{R}^2\setminus\{0\})\otimes\mathrm{L}^2([0,d]),
\end{eqnarray*}
where $\widetilde{\mathbf{A}}:=\alpha\left(
  \frac{y}{x^2+y^2},\frac{-x}{x^2+y^2}\right)$. Consider the quadratic
  form
  \begin{eqnarray}\label{hard}
  \nonumber\widetilde{q}[u]&=&\int_{\mathbb{R}^2} \mid(
  i\nabla+\widetilde{\mathbf{A}})u\mid^2dxdy\\
 \nonumber&=&\int_{\mathbb{R}^2}\left|\left(i \partial_x+\alpha\frac{y}{x^2+y^2}\right)u
  \right|^2dxdy+\int_{\mathbb{R}^2}\left|\left(i
  \partial_y-\alpha\frac{x}{x^2+y^2}\right)u\right|^2dxdy.\\
&&
\end{eqnarray}
By introducing polar coordinates we get
$$r=\sqrt{x^2+y^2};\qquad \frac{x}{r}=\cos\theta,\qquad \frac{y}{r}=\sin\theta,$$
and $$\frac{\partial\theta}{\partial
x}=\frac{-y}{r^2},\quad\frac{\partial\theta}{\partial
y}=\frac{x}{r^2},\quad
\partial_x=\cos\theta\frac{\partial}{\partial
r}-\frac{y}{r^2}\frac{\partial}{\partial\theta},\quad\partial_y=\sin\theta\frac{\partial}{\partial
r}+\frac{x}{r^2}\frac{\partial}{\partial\theta}.
$$
Hence (\ref{hard}) becomes
\begin{eqnarray}\label{119}
 \widetilde{q}[u]&=&\int\left(\mid
\partial_{r}u\mid^2+\frac{1}{r^2}\mid(i\partial_{\theta}u-\alpha u)\mid^2
\right)rdrd\theta.
\end{eqnarray}
Expanding $u$ into Fourier series with respect to $\theta$
$$ u(r,\theta)=\sum_{k=-\infty}^{\infty}u_k(r)\frac{e^{ik\theta}}{\sqrt{2\pi}},$$
enables us to rewrite (\ref{119}) as
\begin{eqnarray*}
   \int\left(\mid\partial_{r}u\mid^2+\frac{1}{r^2}\mid(i\partial_{\theta}u-\alpha u)\mid^2\right)rdrd\theta&\geq&\int\frac{1}{r^2}\mid(i\partial_{\theta}u-\alpha u)\mid^2rdrd\theta  \\
   &\geq& \int\frac{1}{r^2}\left|\sum_{k}(k+\alpha)u_k(r)\frac{e^{ik\theta}}{\sqrt{2\pi}}\right|^2rdrd\theta.
\end{eqnarray*}
Hence
\begin{equation}\label{har}
  \int_{\mathbb{R}^2} \mid(
  i\nabla+\widetilde{\mathbf{A}})u\mid^2dxdy \geq \min_k\mid k+\alpha\mid^2\int\frac{1}{x^2+y^2}\mid u(x,y)
   \mid^2dxdy.
\end{equation}
Here the form in the right hand side is considered on the function
class $\mathrm{H}^1(\mathbb{R}^2)$, obtained by the completion of
the class $\mathcal{C}_0^{\infty}(\mathbb{R}^2\backslash \{0\})$.
Inequality \eqref{har} is the Hardy inequality in two dimensions
with Aharonov-Bohm vector potential \cite{LA}. This yields that
$\displaystyle\sigma\left((i\nabla+\widetilde{\mathbf{A}})^2\right)\subset[0,
+\infty[$.
\begin{rem}
We notice that in \eqref{A}, we take $\alpha\in(0,1)$ which, by
gauge invariance and symmetry, it's equivalent to
$\alpha\in\mathbb{R}\setminus\mathbb{Z}$.
\end{rem}
Since $\displaystyle \sigma(-\Delta)=\sigma_{ess}(-\Delta)=[0,
+\infty[$, then there exists a Weyl sequences
$\displaystyle\{h_n\}_{n=1}^{\infty}$ for the operator $-\Delta$ in
$\mathrm{L}^2(\mathbb{R}^2)$ at $\lambda\geq 0$. Construct the
functions
\begin{equation*}
 \varphi_n(x,y)=\left\{
\begin{array}{ll} h_n & \mbox{if }  x>n \mbox{ and } y>n ,\\ 0 & \mbox{ if not.} \end{array}\right.
\end{equation*}
\begin{eqnarray*}
   \parallel\left((i\nabla+\widetilde{\mathbf{A}})^2-\lambda\right) \varphi_n \parallel&\leq&\parallel(\Delta-\lambda) \varphi_n\parallel+ \parallel
    \widetilde{\mathbf{A}}^2 \varphi_n\parallel+\parallel\widetilde{\mathbf{A}}\nabla\varphi_n\parallel \\
    &\leq& \parallel(\Delta-\lambda)
    \varphi_n\parallel+\frac{c}{n}.
\end{eqnarray*}
Where $c$ is positive real.\\
 Therefore, the functions
$\displaystyle\psi_n=\frac{\varphi_n}{\parallel \varphi_n\parallel}$
is Weyl sequence for
$\displaystyle(i\nabla+\widetilde{\mathbf{A}})^2$ at $\lambda\geq
0$, thus $\displaystyle [0,
+\infty[\subset\sigma_{ess}\left((i\nabla+\widetilde{\mathbf{A}})^2\right)\subset\sigma\left((i\nabla+\widetilde{\mathbf{A}})^2\right)$.\\

Then we get that the spectrum of $\displaystyle
(i\nabla+\widetilde{\mathbf{A}})^2$ is $ \displaystyle [0,+\infty[$,
we know that the spectrum of $\displaystyle -\Delta_{[0,d]}^0$ and
$\displaystyle -\Delta_{[0,d]}^{\infty}$ is $\displaystyle\{
(\frac{j\pi}{d})^{2},\quad j\in\mathbb{N^{\star}}\}$ and
$\displaystyle\{ (\frac{(2j+1)\pi}{2d})^{2},\quad j\in\mathbb{N}\}$
respectively. Therefore we have the spectrum of $H^0$ is
$\displaystyle[(\frac{\pi}{d})^2, +\infty[$. And the spectrum of
$H^{\infty}$ coincides with $\displaystyle[(\frac{\pi}{2d})^2,
+\infty[$.
$\blacksquare$\\

 Consequently, we have
  $$\left[(\frac{\pi}{d})^2, +\infty\right[ \subset \sigma(H)\subset\left[(\frac{\pi}{2d})^2,
  +\infty\right[.$$
Using the property that the essential spectra is preserved under
compact perturbation, we deduce that the essential spectrum of $H$
is $$ \sigma_{ess}(H)=\left[(\frac{\pi}{d})^2, +\infty\right[.$$
\begin{rem}\label{r22}
Let $H^a$ and $H^{a'}$ be two operators defined on \eqref{13}, with
a discs windows of radius $a$ and $a'$ respectively.\\We notice that
$\inf\sigma(H^{a})\leq\inf\sigma(H^{a'})$ if $a\leq a'$.
\end{rem}
\subsection{Preliminary: Cylindrical coordinates}
Let's notice that the system has a cylindrical symmetry. Therefore,
it is natural to consider the cylindrical coordinates system
$(r,\theta,z).$ Indeed, we have that
\begin{eqnarray*}
  \mathrm{L}^2(\Omega, dxdydz)&=&
  \mathrm{L}^2(]0,+\infty[\times[0,2\pi[\times[0,d], rdrd\theta dz).
\end{eqnarray*}
Consequently, a corresponding orthonormal basis in $\mathbb{R}^3$
is given by the three vectors\\
$$e_r=
   \left (
   \begin{array}{ccc}
      \cos\theta \\
      \sin\theta \\
      0  \\
   \end{array}
   \right )
, \qquad e_{\theta}=
   \left (
   \begin{array}{ccc}
      -\sin\theta \\
      \cos\theta \\
      0  \\
   \end{array}
   \right ), \qquad e_z=
   \left (
   \begin{array}{ccc}
      0 \\
      0 \\
      1 \\
   \end{array}
   \right ).$$
We note by $\mathbf{A}_{\theta}$, the Aharonov-Bohm magnetic
potential vector
 \eqref{A} in cylindrical coordinates given by
\begin{equation*}
\mathbf{A}_{\theta}(r,\theta,z)=\frac{\alpha}{r}(-\sin\theta,
\cos\theta, 0)=\frac{\alpha}{r}e_{\theta}.
\end{equation*}
We denote the gradient in cylindrical coordinates by
$\nabla_{r,\theta,z}$. While the operator $i\nabla+\mathbf{A}$ in
cylindrical coordinates is given by
\begin{equation*}
    i\nabla_{r,\theta,z}+\mathbf{A}_{\theta}=i\frac{\partial}{\partial
    r}e_r+\frac{1}{r}\left (i\frac{\partial}{\partial
    \theta}+\alpha\right )e_{\theta}+i\frac{\partial}{\partial
    z}e_z.
\end{equation*}
\noindent Thus the Aharonov-Bohm Laplacian operator in cylindrical
coordinates is given by
\begin{eqnarray*}
  H_{r,\theta,z}&:=&(i\nabla_{r,\theta,z}+\mathbf{A}_{\theta})^2 \\&=&-\frac{1}{r}\frac{\partial}{\partial r}(r\frac{\partial}{\partial
  r})+\frac{1}{r^2}(i\frac{\partial}{\partial\theta}+\alpha)^2 -\frac{\partial^2}{\partial^2
  z}.
\end{eqnarray*}

\section{Results and discussions}
The main result of this paper is the following:
\begin{theorem}\label{thp}
\label{th1} Let $H$ be the operator defined on \eqref{13} and
$\alpha\in (0,1)$. There exist $a_0>0$ such that for any $a$,$d>0$
so that $\displaystyle 0<\frac{a}{d}<a_0$, we have
$$\sigma_d(H)=\emptyset. $$
There exist $a_1>0$ such that for any $a$,$d>0$ so that
$\displaystyle\frac{a}{d}>a_1$, we have
$$\sigma_d(H)\neq\emptyset. $$
\end{theorem}
Before giving the proof of the main result, let us give a series of
remarks.
\begin{rems}
\begin{enumerate}
    \item  The presence of magnetic field in three dimensional straight strip
of width d with the Neumann boundary condition on a disc window of
radius $a$ such that $\displaystyle 0<\frac{a}{d}<a_0$ and Dirichlet
boundary conditions on the remained part of the boundary, destroys
the creation of discrete eigenvalues below the essential spectrum.
If $\displaystyle\frac{a}{d}>a_1$, the effect of the magnetic field
is reduced.
    \item The magnetic destruction of the discrete spectrum
mentioned in this result has the same type as the Hardy-type
inequality for two dimensional quantum waveguide  give in \cite{EK}.
    \item We have to mention that the switching on/off of the
the bound states by the Aharonov-Bohm field for the different
structure was predicted in \cite{Olendski3}.
    \item This result is still true for more general Neumann window containing
some disc. To get the optimal result of $a_0$ and $a_1$, we need
explicit calculation. Let $a'$ is the the inner radii, and $a''$ is
the the outer radii then the conditions will be related to
$\displaystyle \frac{a''}{d}<a_0$ and
$\displaystyle\frac{a'}{d}>a_1$.
\end{enumerate}
\end{rems}

\subsection{Proof of Theorem \ref{thp}}
 As in \cite{n}, let us split $\mathrm{L}^2(\Omega, rdrd\theta dz)$
as follows:
$$\mathrm{L}^2(\Omega, rdrd\theta dz)=\mathrm{L}^2(\Omega_a^-,
rdrd\theta dz)\oplus\mathrm{L}^2(\Omega_a^+, rdrd\theta dz),$$ with
\begin{eqnarray*}
   \Omega_a^-&=& \{(r,\theta, z)\in [0,a]\times[0,2\pi[\times[0,d]\}, \\
\Omega_a^+ &=& \Omega\setminus\Omega_a^-.
\end{eqnarray*}
Therefore
 $$H^{-,N}_a\oplus H^{+,N}_a\leq H \leq H^{-,D}_a\oplus H^{+,D}_a.$$
Here we index by D and N depending on the boundary conditions added
on the surface $r=a$. The min-max principle leads to
\begin{equation}
\sigma_{ess}(H)=\sigma_{ess}( H^{+,N}_a)=\sigma_{ess}(
H^{+,D}_a)=[(\frac{\pi}{d})^2,+\infty[.
\end{equation}
 Let us denote by
$\lambda_k(H^{-,N}_a)$, $ \lambda_k(H^{-,D}_a)$ and $\lambda_k(H)$,
the $k-$th eigenvalue of $H^{-,N}_a$, $H^{-,D}_a$ and $H$
respectively then, again the min-max principle yields the following
\begin{equation}\label{eqv}
\lambda_k(H^{-,N}_a) \leq\lambda_k(H)\leq\lambda_k(H^{-,D}_a),
\end{equation}
and for $k\geq 2$
\begin{equation}
\lambda_{k-1}(H^{-,N}_a) \leq\lambda_k(H)\leq\lambda_k(H^{-,N}_a);
\end{equation}
\begin{equation}\label{direl}
\lambda_{k-1}(H^{-,D}_a) \leq\lambda_k(H)\leq\lambda_k(H^{-,D}_a).
\end{equation}
Thus, if $H^{-,N}_a$ does not have a discrete spectrum below
$(\frac{\pi}{d})^2$, then $H$ do not has as well. We mention that
it's a sufficient condition. \\The eigenvalue equation is given by
\begin{equation}\label{eq}
  H^{-,N}_af(r,\theta,z) = \lambda f(r,\theta,z).
\end{equation}
This equation is solved by separating variables and considering
$\displaystyle f(r,\theta,z)=R(r)P(\theta)Z(z)$.\\ We divide the
equation \eqref{eq} by $f$, we obtain
\begin{equation}\label{eeq}
 \frac{1}{R}(R''+\frac{1}{r}R')+\frac{1}{r^2}\frac{1}{P}(i\frac{\partial}{\partial\theta}+\alpha)^2P+\frac{Z''}{Z} = \lambda.
\end{equation}
Plugging the last expression in equation \eqref{eeq} and first
separate $Z$ by putting all the $z$ dependence in one term so that
$\displaystyle\frac{Z''}{Z}$ can only be constant. The constant is
taken as $\displaystyle -k_z^2=-(\frac{(2j+1)\pi}{2d})^2$ for
convenience.\\ Second, we separate the term $ \displaystyle
\frac{1}{P}(i\frac{\partial}{\partial\theta}+\alpha)^2P$ which has
all the $\theta$ dependance. Using the fact that the problem has an
axial symmetry and the solution has to be $2\pi$ periodic and single
value in $\theta$, we obtain $\displaystyle
\frac{1}{P}(i\frac{\partial}{\partial\theta}+\alpha)^2P$ should be a
constant $\displaystyle -(m-\alpha)^2=-\nu^2$ for
$m\in\mathbb{Z}$.\\
Finally, we get the following equation \eqref{eeq} for $R$
\begin{equation}\label{eq1}
R''(r)+\frac{1}{r}R'(r)+[\lambda-k_z^2-\frac{\nu^2}{r^2}]R(r)=0.
\end{equation}
We notice that the equation \eqref{eq1}, is the Bessel equation and
its solutions could be expressed in terms of Bessel functions. More
explicit solutions could be given by considering boundary
conditions.\\
 The solution of the equation \eqref{eq1} is given by $
\displaystyle R(r)=cJ_{\nu}(\beta r)$, where $\displaystyle c\in
\mathbb{R}^{\star}$, $\displaystyle \beta^2=\lambda-k_z^2$ and
$J_{\nu}$ is the
Bessel function of first kind of order $\nu$. \\
We assume that
\begin{eqnarray}\label{nom}
  \nonumber R'(a)=0 &\Leftrightarrow& J'_{\nu}(\beta a)=0 \\
   &\Leftrightarrow& a\beta=x'_{\nu,n}.
\end{eqnarray}
 Where $x'_{\nu,n}$ is the $n-$th positive zero of the Bessel function
 $J'_{\nu}$.\\
 Consequently to equation (\ref{nom}), $H^{-,N}_a$ has a sequence of
eigenvalues given by
\begin{eqnarray*}
    \lambda_{j,\nu,n} &=& \frac{x_{\nu,n}'^2}{a^2}+k_z^2 \\
      &=&\frac{x_{\nu,n}'^2}{a^2}+\left(\frac{(2j+1)\pi}{2d}\right)^2.
   \end{eqnarray*}
As we are interested for discrete eigenvalues which belongs to
$\displaystyle[\left(\frac{\pi}{2d}\right)^2,\left(\frac{\pi}{d}\right)^2)$
only $\lambda_{0,\nu,n}$ intervenes.\\
 If
 \begin{equation}\label{cond1}
\left(\frac{\pi}{d}\right)^2\leq\lambda_{0,\nu,n},
 \end{equation}
 then there $H$ does not have a discrete spectrum.
We recall that $\nu^2=(m-\alpha)^2$ and it is related to magnetic
flux, also recall that $x'_{\nu,n}$ are the positive zeros of the
Bessel function
 $J'_n$ and $\forall\nu>0$, $\forall n\in\mathbb{N}^{\star}$; $0<x'_{\nu,n}<x'_{\nu,n+1}$ (see \cite{SF}).
 So, for any eigenvalue of $ H^{-,N}_a$,
 $$\frac{x_{\nu,1}'^2}{a^2}+\left(\frac{\pi}{2d}\right)^2<\frac{x_{\nu,n}'^2}{a^2}+\left(\frac{\pi}{2d}\right)^2=\lambda_{0,\nu,n}.$$
 An immediate consequence of the last inequality is that to satisfy \eqref{cond1} it is
sufficient to have
$$3\left(\frac{\pi}{2d}\right)^2< \frac{x'^2_{\nu,1}}{a^2},$$
therefore
$$\frac{\sqrt{3}\pi}{2d}< \frac{x'_{\nu,1}}{a},$$
then
$$\frac{a}{d}< \frac{2 x'_{\nu,1}}{\sqrt{3}\pi} .$$
We have (see \cite{LGC,LL,RP,SF})
\begin{equation*}
    \nu+\alpha_n \nu^{1/3}<x'_{\nu,n},
\end{equation*}
where $\alpha_n=2^{-1/3}\beta_n$ and $\beta_n$ is the $n-$th
positive root of the equation
\begin{equation*}
    J_{\frac{2}{3}}\left( \frac{2}{3} x^{3/2}\right)-J_{\frac{-2}{3}}\left( \frac{2}{3}
    x^{3/2}\right)=0.
\end{equation*}
For $n=1$, we have $\displaystyle\alpha_n \nu^{1/3}\approx 0.6538$
(see \cite{LGC}), then
\begin{equation}\label{bessel}
 c_0:=0.6538+\alpha <0.6538+\nu<x'_{\nu,1}.
 \end{equation} Then we get that for $d$ and $a$
positives such that $\displaystyle
\frac{a}{d}<a_0:=\frac{2c_0}{\sqrt{3}\pi}$,
$$\sigma_d(H)=\emptyset. $$
This ends the proof of the first result of the theorem \ref{thp}.

 By the min-max principle and \eqref{eqv}, we know that if $H^{-,D}_a$ exhibits a
discrete spectrum below $(\frac{\pi}{d})^2$, then $H$ do as well.\\
$H^{-,D}_a$ has a sequence of eigenvalues \cite{n,wat}, given by
\begin{eqnarray*}
    \lambda_{j,\nu,n} &=&  \left(\frac{x_{\nu,n}}{a}\right)^2+\left(\frac{(2j+1)\pi}{2d}\right)^2.\\
   \end{eqnarray*}
 Where $x_{\nu,n}$ is the $n-$th positive zero of Bessel function of order
 $\nu$ ( see \cite{n}). As we are interested for discrete eigenvalues which belongs to
$\displaystyle[(\frac{\pi}{2d})^2,(\frac{\pi}{d})^2)$ only for
$\lambda_{0,\nu,n}$.\\
If the following condition
 \begin{equation}\label{cond}
\lambda_{0,\nu,n}<\left(\frac{\pi}{d}\right)^2
 \end{equation}
 is satisfied, then $H$ have a discrete spectrum.\\
 We recall that $0<x_{\nu,n}<x_{\nu,n+1}$ for any $\nu>0$ and any
 $n\in\mathbb{N}^{\star}$ (see \cite{SF}). So, for any eigenvalue of
 $H^{-,D}_a$,
 $$\frac{x_{\nu,1}^2}{a^2}+\left(\frac{\pi}{2d}\right)^2< \frac{x_{\nu,n}^2}{a^2}+\left(\frac{\pi}{2d}\right)^2=\lambda_{0,\nu,n}.$$
 An immediate consequence of the last inequality
is that to satisfy \eqref{cond} it is sufficient to set then
$$\frac{2 x_{\nu,1}}{\sqrt{3}\pi}< \frac{a}{d}.$$  We have (see \cite{LGC,SF})
\begin{equation*}
    \displaystyle \sqrt{\left(n-\frac{1}{4}\right)^2\pi^2+\nu^2}<x_{\nu,n}.
\end{equation*}
For $n=1$, we have
\begin{equation}\label{bessel}
 c_1:=\sqrt{\left(\frac{3\pi}{4}\right)^2+\alpha^2}<\sqrt{\left(\frac{3\pi}{4}\right)^2+\nu^2}<x_{\nu,1}.
 \end{equation} Then we get that for $d$ and $a$
positives such that $\displaystyle
\frac{a}{d}>a_1:=\frac{2c_1}{\sqrt{3} \pi}$,
$$\sigma_d(H)\neq\emptyset. $$
\begin{flushright}
$\blacksquare$
\end{flushright}

\section{Numerical computations}
This section is devoted to some numerical computations. We represent
the radius values $\displaystyle
a'_0:\alpha\mapsto\frac{2x'_{\nu,1}}{\sqrt{3}\pi}$ and
$\displaystyle a'_1:\alpha\mapsto\frac{2x_{\nu,1}}{\sqrt{3}\pi}$,
where $\displaystyle x'_{\nu,1}$ and $\displaystyle x_{\nu,1}$ is
the first positive zero of Bessel function $\displaystyle J'_{\nu}$
and $\displaystyle J_{\nu}$ respectively, a function of the flux
magnetic $\displaystyle \alpha$, which makes it possible to exist or
not discrete eigenvalues of $H$.
\newpage
\begin{figure}
\centering
\includegraphics[width=1\textwidth]{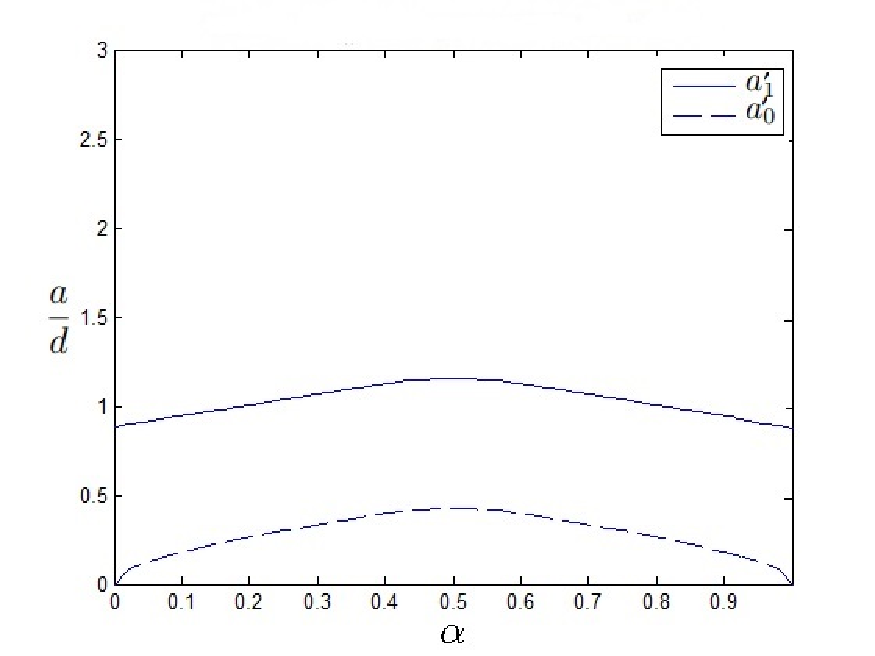}
\caption{\label{fig2}The areas of the existence and the absence of
eigenvalues of $H$.}
\end{figure}

We observe that the result is invariant with respect to the
transformation $\displaystyle \alpha\rightarrow 1-\alpha$ as the
Hardy-type inequality (\ref{har}) suggests.\\ The area below the
dashed curve corresponds to the absence of the eigenvalues of $H$
and the area above the curve of $a'_1$ corresponds to their
existence. But we can not conclude the existence or absence of
eigenvalues of $H$ in the area between the two curves.

Using the min-max principle and the monotonicity argument of Remark
\ref{r22}, we can say that the critical radius as a function of
$\alpha$ exists, its is a continuous function having a curve laying
between the two curves. Its is symmetric with respect to
$\displaystyle\alpha=\frac{1}{2}$, and having zero value at the
origin.


We set the flux magnetic $\displaystyle\alpha=0.5$,  $\displaystyle
a=0.3$ in figure \ref{fig4} such that $ \displaystyle \frac{a}{d}$
between the two critical values $\displaystyle  a_0$ and
$\displaystyle a_1$. We represent 
$\displaystyle  d\mapsto(\frac{\pi}{2d})^2+(\frac{x'(0.5)}{a})^2$
where $\displaystyle
 x'(0.5)$ is the first zero of the bessel functions $\displaystyle
 J'_{0.5}$, according to $\displaystyle  a$.
\newpage
\begin{figure}[h] \centering
\includegraphics[width=1\columnwidth]{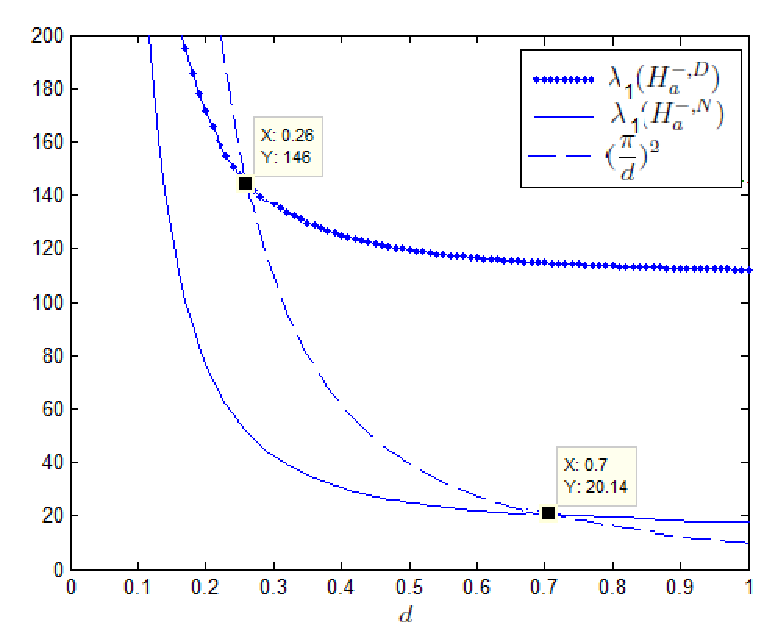}
\label{fig4}
 \caption{\label{fig4} $d\mapsto (\frac{\pi}{2d})^2+(\frac{x^i(0.5)}{a})^2$ where
 $x^0(0.5)$ and $x^1(0.5)$ is the first zero of the bessel function $J_{0.5}$ and $J'_{0.5}$
respectively.}
\end{figure}

 {\bf{Acknowledgements:}}
The authors are grateful to Philippe BRIET for useful remarks and
suggestions and for the hospitality of the two authors in CPT. We
would like to thank the unknown referees for their valuable comments
and suggestions wish improve the paper.

\end{document}